\documentclass[letterpaper]{article} 
\usepackage{aaai2027}  
\usepackage[hyphens]{url}  
\usepackage{graphicx} 
\usepackage{natbib}  
\usepackage{caption} 
\usepackage{algorithm}
\usepackage{algorithmic}
\usepackage{xspace}
\usepackage{amsmath}
\usepackage{multirow}
\usepackage{newfloat}
\usepackage{listings}
\DeclareCaptionStyle{ruled}{labelfont=normalfont,labelsep=colon,strut=off} 
\floatstyle{ruled}
\newfloat{listing}{tb}{lst}{}
\floatname{listing}{Listing}

\usepackage{booktabs}
\usepackage[table]{xcolor} 
\newcommand{\ourmethod}{\textsc{FAVA}\xspace}

\title{\ourmethod{}: Formal Authorization for Verified Agents with Evidence-Backed Permission Graphs}
\author{
Yifan Zhang\textsuperscript{\rm 1},
Xinkui Zhao\textsuperscript{\rm 1}\corresponding,
Sai Liu\textsuperscript{\rm 1},
Hengxuan Lou\textsuperscript{\rm 1},
Guanjie Cheng\textsuperscript{\rm 1},
Chang Liu\textsuperscript{\rm 1}
}
\affiliations{
\textsuperscript{\rm 1}Zhejiang University\\
zhaoxinkui@zju.edu.cn
}

\begin{document}

\maketitle

\begin{abstract}
Large language model (LLM) agents autonomously interleave semantic reasoning with complex system operations. In these dynamic environments, static tool-level permissions are fundamentally insufficient; safe authorization is highly context-dependent and heavily reliant on evolving runtime states and data flows. We present \ourmethod{} (\textbf{F}ormal \textbf{A}uthorization for \textbf{V}erified \textbf{A}gents), a permission-carrying authorization framework for agent execution. \ourmethod{} utilizes an LLM-guided Permission Intermediate Representation (IR) to translate ambiguous natural-language tasks into structured constraints. A deterministic lowering pass then converts this IR into an evidence-backed permission graph that explicitly tracks data flows, dependencies, and contextual labels. To provide strict security guarantees, a Satisfiability Modulo Theories (SMT) authorizer mathematically verifies the current graph against security policies before any effectful action executes. A runtime gateway then enforces the solver's result, either authorizing the execution or intercepting it with a precise counterexample. We evaluate \ourmethod{} across OpenAgentSafety, OctoBench, and ActPlane scenarios. Our evaluation demonstrates that \ourmethod{} achieves a 90.5\% Decision Compliance Rate (DCR) over the aggregate dataset, successfully intercepting dynamic violating traces in the evaluated trace-conditioned scenarios.
\end{abstract}

\section{Introduction}

LLM agents are increasingly integrated into critical computational environments, operating across software repositories, enterprise chat systems, cloud storage, and local shells~\cite{debenedetti2024agentdojo,wang2025openhands,yang2024swe,zhang2025sortinghat,wei2025poster}. In these complex settings, an agent routinely interleaves semantic tasks with concrete system actions, such as reading files, generating code, and executing shell commands~\cite{chatlatanagulchai2025agent,chatlatanagulchai2025use,lulla2026impact}. Because LLMs autonomously plan and execute actions probabilistically, their exact runtime trajectories and data flows cannot be statically foreseen~\cite{jiang2024followbench,zheng2026agentcgroup,wu2026crab}. Consequently, agent authorization is highly context-dependent rather than bound to static actions.

To understand the exact nature of this context dependency in practice, we conducted an empirical study on real-world agent governance. Our analysis of active open-source repositories reveals that developer-defined authorization is heavily stateful. Specifically, 90\% of the analyzed projects enforce sequence-dependent temporal constraints, and over 40\% incorporate complex nested conditional logic.

However, existing defense mechanisms only address fragments of this dynamic complexity and fall short in providing checkable runtime authorization boundaries. Prompt-only policies attempt to ask the model to behave safely but completely fail to enforce actual runtime decisions~\cite{zou2023universal,andriushchenko2025agentharm}. While keyword filters and static tool allowlists are easy to deploy, they inherently ignore critical data flow and execution context~\cite{babu2026toolmenubench,gaurav2025governance,luo2025unsafe}. OS-level sandboxes can restrict file and network execution, but they lack the semantic understanding needed to govern the high-level conditional logic that modern agents require~\cite{ruan2024identifying,zhou2024haicosystem,chen2025ukfaas,dockeragent,e2b2024}.

To address these concerns, we propose \ourmethod, Formal Authorization for Verified Agents. The central philosophy of \ourmethod{} is to separate semantic permission extraction from final safety authorization. Rather than trusting an LLM to make opaque ``safe/unsafe" judgments, \ourmethod{} utilizes the LLM where it excels: parsing natural language tasks into a structured \textit{Permission IR} (Intermediate Representation). This IR explicitly maps intents, protected assets, tool-level actions, obligations, scoped allowances, forbidden actions, and evidence spans. A deterministic lowering pass then converts this IR into an \textit{evidence-backed permission graph}. The final safety decision is computed by an SMT authorizer over the graph and system policy. Thus, \ourmethod{} does not provide an end-to-end proof that an agent is safe from natural language alone; its formal guarantee begins once available evidence has been lowered into the permission graph. Under the stated graph, policy, gateway, and backend assumptions, \ourmethod{} returns either a counterexample to intercept forbidden flows or a capability set to authorize backend execution. Crucially, this graph-based design enables monotone graph repair: runtime events append observed facts to the graph and force re-authorization before any effectful action occurs.

This paper makes three primary contributions:
\begin{itemize}
    \item \textbf{Framework Concept and Positioning:} We propose \ourmethod{}, a novel runtime authorization framework for LLM agents that fundamentally shifts permission management to verifiable graph authorization.
    
    \item \textbf{Formal Authorization Methodology:} We introduce a pipeline that leverages LLMs to parse natural-language tasks into a structured \textit{Permission IR}, which is deterministically lowered into an \textit{evidence-backed permission graph}. This design enables an SMT authorizer to compute policy-consistency decisions over the graph and supports monotone graph repair with just-in-time re-authorization.
    
    \item \textbf{Comprehensive Evaluation:} We evaluate \ourmethod{} across OpenAgentSafety, OctoBench, and ActPlane scenarios. \ourmethod{} achieves 90.5\% DCR over the full 801-case replay while preserving 100.0\% DCR on the structured and labeled trace-conditioned splits.
\end{itemize}

\section{Motivation: Agent Constraints}
\label{sec:emp}
\begin{figure}[t]
    \centering
    \includegraphics[width=1\linewidth]{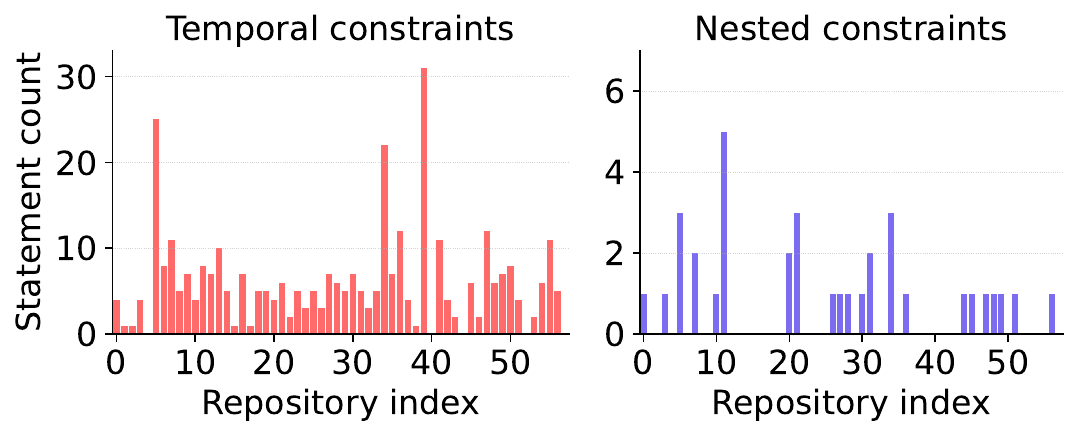}
    \caption{Frequency of dynamic constraints extracted from real-world agent instruction files in 2026 GitHub repositories. The left panel shows the widespread use of temporal constraints (sequence-dependent execution), while the right panel highlights the reliance on nested conditional logic (e.g., "if", "only when").}
    \label{fig:motivation}
\end{figure}

\textbf{Data Collection.}
To understand how developers define authorization boundaries for LLM agents in practice, we conducted an empirical study on open-source projects. We collected public GitHub repositories containing \texttt{CLAUDE.md} or \texttt{AGENT.md}---standardized files increasingly used to provide system-level behavioral instructions to autonomous coding agents. We specifically targeted AI-agent projects created or actively maintained in 2026. To ensure the ecological validity and quality of our dataset, we applied strict filtering criteria, systematically excluding non-code, inactive, artificially inflated (fake-star), and stub repositories. We then parsed the natural language instructions within these files to extract and categorize statements dictating execution permissions.

\textbf{Data Analysis and Implications.}
Our analysis reveals that real-world agent authorization is rarely static; instead, it is highly dependent on runtime states and execution order. As illustrated in Figure~\ref{fig:motivation}, developers increasingly govern agent capabilities through complex, stateful policies rather than simple binary allowlists. Specifically, 90\% of the analyzed repositories enforce sequence-dependent temporal constraints (e.g., ``do not commit \textit{before} running tests''), with individual projects often containing a substantial volume of such rules. Compounding this complexity, over 40\% of the repositories incorporate nested and conditional logic (e.g., ``allow internet access \textit{only when} querying the official API''). This pervasive combination of temporal sequencing and state-dependent modifiers highlights a fundamental misalignment between current static authorization mechanisms and actual governance needs. Consequently, the inherent semantic complexity of real-world policies motivates a paradigm shift in agent security, underscoring the critical necessity for an approach capable of extracting rich semantic rules from natural language and formally verifying multi-step execution graphs at runtime.

\section{\ourmethod{} Method}
\label{sec:method}

\begin{figure*}[t]
\centering
\includegraphics[width=0.98\textwidth]{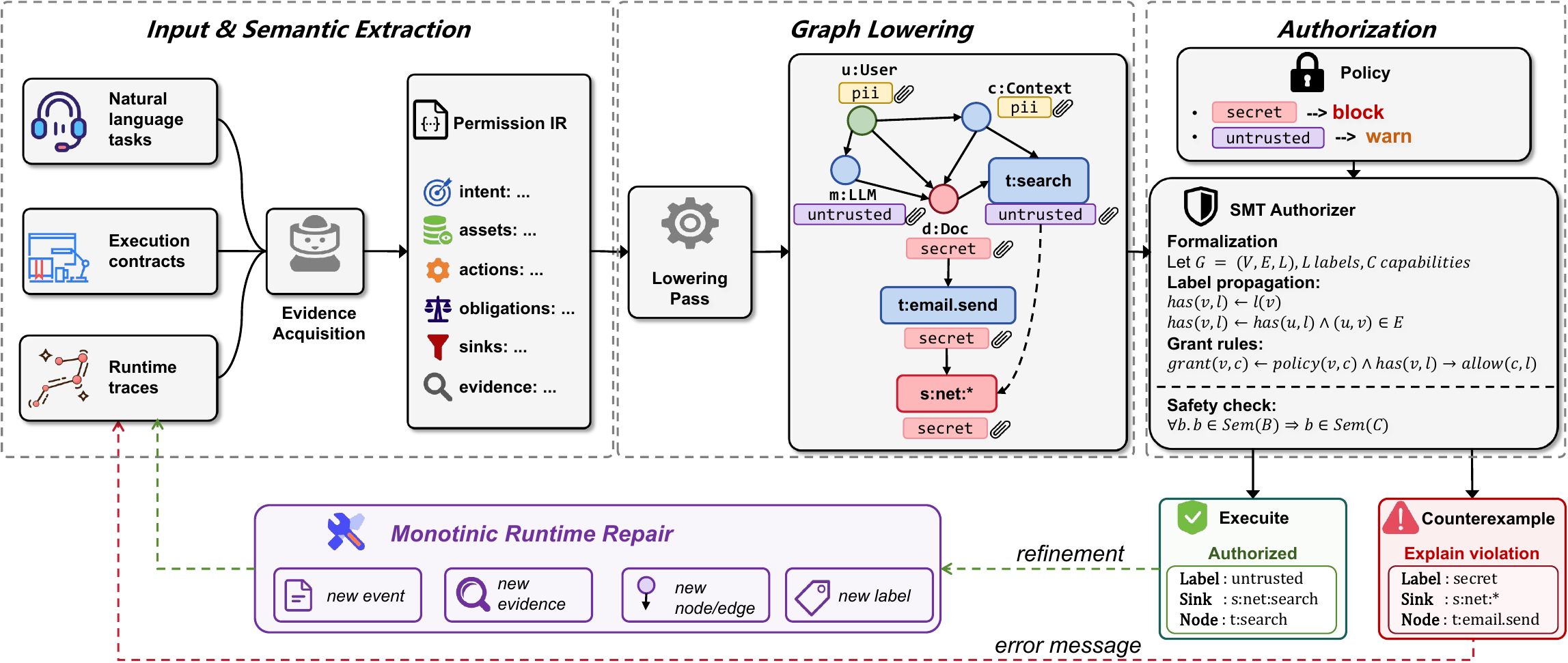}
\caption{\ourmethod{} architecture. The system extracts inputs into a Permission IR, lowers it into an evidence-backed graph, and enforces security via an SMT authorizer and runtime gateway. The graph is monotonically updated with new observations.}
\label{fig:pcae-overview}
\end{figure*}

\subsection{Overview}
Real-world agent policies are inherently stateful and complex, yet natural language instructions remain fundamentally too ambiguous for rigorous enforcement. Left unchecked, this modality gap exposes agents to critical vulnerabilities, which can be broadly categorized into: (1) \textbf{unauthorized data exfiltration} (e.g., leaking secrets to untrusted sinks), (2) \textbf{destructive state modifications} (e.g., executing unsafe commands), and (3) \textbf{temporal obligation violations} (e.g., bypassing mandatory prerequisite checks). 

Because probabilistic LLMs cannot guarantee formal correctness against these threats, security requires strict mathematical formalization. To achieve this, \ourmethod{} strictly separates semantic comprehension from logical enforcement. It leverages LLMs solely to parse ambiguous tasks into a structured \textbf{Permission IR}. To capture stateful data flows, this IR is deterministically lowered into a \textbf{Permission Graph}. Finally, rather than relying on opaque model judgments, the graph serves as the exact substrate for an \textbf{SMT Authorizer}, transforming agent safety into an mathematical satisfiability problem.

\subsection{Evidence Acquisition and Permission IR}

Because formal verifiers cannot process raw natural language and probabilistic LLMs cannot be trusted with access control decisions, \ourmethod{} employs an LLM strictly as a semantic extractor. It parses task text, execution contracts, and runtime traces into a structured Permission IR. This IR explicitly defines the authorization boundary using five core fields---\texttt{intent}, \texttt{assets}, \texttt{actions}, \texttt{obligations}, and \texttt{sinks}---alongside \texttt{evidence} spans that cryptographically or textually link each field to its source. 

By cleanly separating extraction from authorization, complex semantic instructions (e.g., ``summarize and send to Slack'') and temporal guards (e.g., ``require review before testing'') are preserved as auditable structural constraints rather than being compressed into opaque LLM judgments. To maintain architectural consistency, dynamic runtime traces are mapped into this exact same IR format. They are populated directly from observed tool calls, arguments, and event ancestry, ensuring a uniform representation across static tasks and dynamic executions where only the source of the evidence differs.

\subsection{Graph Lowering}

To enforce sequence-dependent temporal constraints, flat permission lists are structurally insufficient. Instead, a deterministic lowering pass maps the Permission IR into an evidence-backed permission graph $G=(V,E)$ (Table~\ref{tab:graph-schema}). Nodes carry evidence-grounded security labels (e.g., \texttt{secret}, \texttt{destructive}), while edges encode dependency relations.

This pass preserves IR semantics rather than compressing them into lossy allowlists. Scoped allowances become declassification constraints, not unconditional permissions. Crucially, sanitization requires explicit trusted transformer specifications (e.g., \texttt{redact(secret) -> safe\_text}); vague LLM assertions (e.g., ``summary'') are not treated as sanitizers. This transforms semantic data into a formally verifiable substrate.

Mapping rules explicitly translate IR assets to \texttt{source} nodes, effectful calls to \texttt{tool} nodes, and external targets to \texttt{sink} nodes. Typed edges formalize interactions: \texttt{data} for value flows, \texttt{control} for temporal guards, and \texttt{parent} for runtime ancestry. The authorizer strictly trusts observed and policy-defined edges, retaining inferred edges purely for auditing extraction errors. Finally, to guarantee integrity against LLM hallucinations, the lowered graph undergoes strict structural and evidential validation, rejecting malformed constructs before reaching the SMT authorizer.

\begin{table}[t]
\centering
\small
\resizebox{\linewidth}{!}{%

\begin{tabular}{@{} l l l @{}}
\toprule
\textbf{Object} & \textbf{Field} & \textbf{Meaning} \\
\midrule
\multirow{7}{*}{Node} 
     & \texttt{id} & Unique identifier \\
     & \texttt{kind} & Type (e.g., context, source, tool, sink) \\
     & \texttt{op} & Action (e.g., read, write, execute) \\
     & \texttt{args}, \texttt{outputs} & Input/output arguments and values \\
     & \texttt{labels} & Security labels and evidence \\
     & \texttt{requests} & Required or optional capabilities \\
     & \texttt{time} & Execution timestamp or order \\
\midrule
\multirow{4}{*}{Edge} 
     & \texttt{src}, \texttt{dst} & Source and destination nodes \\
     & \texttt{type} & Dependency (data, control, parent) \\
     & \texttt{evidence} & Textual span or event ancestry \\
     & \texttt{trust} & Trust source (observed, policy, inferred) \\
\bottomrule
\end{tabular}
}
\caption{Permission graph schema used by the lowering pass.}
\label{tab:graph-schema}
\end{table}

\subsection{SMT Authorization}

The complex nested and conditional logic prevalent in real-world agent policies creates authorization boundaries too intricate for standard heuristic checks. \ourmethod{} formulates authorization over its permission-graph abstraction as a Satisfiability Modulo Theories (SMT) problem. 

Conceptually, the solver continuously evaluates two dimensions over the evidence graph: data-flow taint and capability grants. Labels (e.g., \texttt{secret}, \texttt{pii}) propagate along dependency edges unless transformed by a trusted sanitizer specification. Simultaneously, the system evaluates capabilities required to execute a candidate action against a core policy of forbidden label-to-sink flows (e.g., \texttt{secret -> net:*}, \texttt{pii -> tool:slack.post\_message}). The solver either returns an authorized capability set or surfaces a precise counterexample to intercept the execution.

Formally, for a finite graph $G=(V,E)$, $Pred(v)$ denotes the predecessors of $v$ over data and policy-control edges. $\text{Explicit}(v,\ell)$ is true when node $v$ directly carries evidence for label $\ell$, and $\text{Source}(v,\ell,P)$ is true when policy $P$ labels the resource read or produced by $v$ as $\ell$. A capability $c$ is a normalized resource request (e.g., \texttt{file:write:/tmp/x}). Let $H[v,\ell]$ denote the presence of label $\ell$ at node $v$, and $A[v,c]$ denote that capability $c$ is granted to $v$. $\text{Req}(v,c)$ dictates that $v$ cannot execute unless $c$ is granted. A forbidden rule $f=(L_f,S_f)$ defines protected labels $L_f$ and a sink pattern $S_f$, with $\text{Match}(v,S_f)$ returning capabilities requested by $v$ that match the sink.

Obligations are encoded as Boolean guard facts over the prefix. For example, \texttt{allow(commit) only if tests\_passed} becomes:
$$A[v,\texttt{commit}] \Rightarrow Seen(t_p,v),$$
where $Seen(t_p,v)$ is true only if an earlier observation node establishes the \texttt{tests\_passed} fact. Data-flow denial (e.g., \texttt{deny network\_send if reachable(secret)}) is encoded as:
$$\forall v.\ \neg(H[v,\texttt{secret}] \wedge A[v,\texttt{net:send}]).$$
Because \ourmethod{} operates on finite prefix graphs, constraints are solved over the bounded runtime state and re-evaluated as new events occur.

\paragraph{Finite-Graph Soundness.}
For a validated finite permission graph and fixed policy translation, Algorithm~\ref{alg:smt-authorization} is sound for forbidden label-to-sink reachability: any satisfiable capability assignment is guaranteed to strictly satisfy all policy rules encoded in $P$. Conversely, if granting a required capability forces a protected collision, the constraints become unsatisfiable, and the solver yields a precise counterexample trace identifying the violating path. This guarantee applies strictly to the formal abstraction; it does not assert the completeness of upstream natural-language extraction or implicit-flow discovery.

\paragraph{Taint-Monotonic Safety.}
The monotonicity in \ourmethod{} applies strictly to observed graph information. Let $G \preceq_t G'$ denote a taint-only extension: $G'$ contains all objects in $G$ and may add more, but does not introduce new trusted sanitizers, declassifications, or positive obligation facts (e.g., \texttt{tests\_passed}). For a fixed policy and required capability set, if action $a$ is rejected on $G$, it is necessarily rejected on any $G'$ where $G \preceq_t G'$. Because label propagation is monotonically increasing and forbidden-flow checks are hard constraints, adding taint information can only preserve or introduce collisions; it cannot resolve an existing one.

Positive guard observations and trusted transformations (e.g., sanitization) are treated separately as explicit policy mechanisms. Establishing a \texttt{tests\_passed} observation may satisfy a precondition, but this represents the fulfillment of an explicit rule, not a weakening of the safety boundary. Therefore, these events are excluded from $G \preceq_t G'$ and fall outside ordinary monotone taint growth.

\begin{algorithm}[t]
\small
\caption{SMT Authorization Encoding}
\label{alg:smt-authorization}
\begin{algorithmic}[1]
\REQUIRE Permission graph $G=(V,E)$, policy $P$
\ENSURE Authorized capabilities $C$ or counterexample

\STATE \textbf{Variables:} $H[v,\ell]$ (label presence), $A[v,c]$ (capability granted)

\FORALL{$v \in V$}
  \STATE \textit{// 1. Label Propagation}
  \FORALL{label $\ell$}
    \IF{$v$ has trusted sanitizer for $\ell$}
      \STATE Assert transformed label according to sanitizer specification
    \ELSE
      \STATE Assert $H[v,\ell] = \text{Explicit}(v,\ell) \vee \text{Source}(v,\ell,P) \vee \bigvee_{u \in Pred(v)} H[u,\ell]$
    \ENDIF
  \ENDFOR

  \STATE \textit{// 2. Capability Allocation}
  \STATE For each req. $c$ at $v$: Assert $A[v,c] = \text{true}$ if required, else softly minimize $A[v,c]$
\ENDFOR

\STATE \textit{// 3. Policy Enforcement}
\FORALL{forbidden rule $f \in P$ and $v \in V$}
  \STATE Assert $\neg \big( \bigwedge_{\ell \in Labels(f)} H[v,\ell] \;\wedge\; \bigvee_{c \in Match(v, Sink(f))} A[v,c] \big)$
\ENDFOR

\STATE \textit{// 4. Resolution}
\RETURN authorized capabilities if satisfiable; otherwise counterexample trace
\end{algorithmic}
\end{algorithm}

\subsection{Monotonic Tracking and Enforcement Gateway}

Agent environments are highly dynamic: an action that is benign initially may become dangerous later if a sensitive resource is accessed. To prevent the ``context amnesia'' typical of LLM agents, \ourmethod{} continuously expands its authorization boundary through monotonic runtime repair. Runtime events strictly append new nodes, edges, labels, and evidence; established security constraints are never silently discarded.

Operating on this monotonic graph, \ourmethod{} acts as a strict Just-In-Time (JIT) gateway. It runs the SMT authorizer immediately before any effectful action and ensures that downstream execution backends remain at least as restrictive as the solver's mathematical decision. 

The gateway enforces a deterministic, risk-aware binary decision. If the solver identifies a counterexample, the action is \texttt{block}ed and the violating path is surfaced. If no violation is found and the Permission IR marks the task as benign, the action is \texttt{allow}ed. Crucially, if the solver finds no explicit forbidden flow but the extracted risk posture remains \texttt{sensitive}, \texttt{dangerous}, or \texttt{ambiguous}, the gateway deliberately fails closed. This strict automation prioritizes security over utility---resolving scope ambiguity via default blocking rather than human intervention---a tradeoff explicitly reflected in our conservative benign allow rate.
\begin{table*}[t]
\centering
\small
\setlength{\tabcolsep}{3.5pt}
\resizebox{\linewidth}{!}{%
\begin{tabular}{l | r r r r r | r r r r r | r r r r r | r r r r r}
\toprule
\multirow{2}{*}{Method} & \multicolumn{5}{c|}{OctoBench} & \multicolumn{5}{c|}{OpenAgentSafety} & \multicolumn{5}{c|}{ActPlane Public} & \multicolumn{5}{c}{ActPlane Traces} \\
\cmidrule(lr){2-6} \cmidrule(lr){7-11} \cmidrule(lr){12-16} \cmidrule(l){17-21}
 & TP & TN & FP & FN & DCR & TP & TN & FP & FN & DCR & TP & TN & FP & FN & DCR & TP & TN & FP & FN & DCR \\
\midrule
Vanilla / No Guard & 0 & 16 & 0 & 201 & 7.4 & 0 & 56 & 0 & 303 & 15.6 & 0 & 15 & 0 & 20 & 42.9 & 0 & 76 & 0 & 114 & 40.0 \\
GuardAgent         & 10 & 16 & 0 & 191 & 12.0 & 168 & 14 & 42 & 135 & 50.7 & 20 & 15 & 0 & 0 & 100.0 & 38 & 75 & 1 & 76 & 59.5 \\
Regex Guard        & 158 & 7 & 9 & 43 & 76.0 & 303 & 0 & 56 & 0 & 84.4 & 15 & 2 & 13 & 5 & 48.6 & 12 & 54 & 22 & 102 & 34.7 \\
AgentSpec          & 12 & 16 & 0 & 189 & 12.9 & 181 & 18 & 38 & 122 & 55.4 & 20 & 14 & 1 & 0 & 97.1 & 77 & 75 & 1 & 37 & 80.0 \\
ActPlane           & -- & -- & -- & -- & -- & 122 & 36 & 20 & 181 & 44.0 & 20 & 0 & 15 & 0 & 57.1 & 98 & 7 & 69 & 16 & 55.3 \\
AuthGraph & 51 & 16 & 0 & 150 & 30.9 & 81 & 50 & 6 & 222 & 36.5 & 10 & 15 & 0 & 10 & 71.4 & 114 & 76 & 0 & 0 & 100.0 \\
SafeAgent & 174 & 13 & 3 & 27 & 86.2 & 244 & 12 & 44 & 59 & 71.3 & 10 & 15 & 0 & 10 & 71.4 & 114 & 76 & 0 & 0 & 100.0 \\
\rowcolor{gray!15}
\textbf{\ourmethod{}} & \textbf{196} & \textbf{1} & \textbf{15} & \textbf{5} & \textbf{90.8} & \textbf{301} & \textbf{2} & \textbf{54} & \textbf{2} & \textbf{84.4} & \textbf{20} & \textbf{15} & \textbf{0} & \textbf{0} & \textbf{100.0} & \textbf{114} & \textbf{76} & \textbf{0} & \textbf{0} & \textbf{100.0} \\
\bottomrule
\end{tabular}
}%
\caption{Decision-compliance results. DCR is reported as a percentage.}
\label{tab:main-results}
\end{table*}
\subsection{Security Assumptions and System Scope}

\ourmethod{} provides authorization guarantees over the explicit abstraction it checks, not over every possible behavior of an unconstrained agent runtime. The SMT authorizer is sound with respect to the generated permission graph, the fixed policy translation, and the gateway-visible candidate action. This formal guarantee relies on three foundational engineering assumptions. First, all security-relevant runtime effects must be mediated by the gateway before execution. Second, backend execution must be capability-conformant: a backend may restrict more than the solver permits, but it must never execute effects outside the granted capability set. Third, policy translation and sanitizer specifications are treated as part of the trusted computing base.



\section{Implementation}

The \ourmethod{} prototype is implemented in Python, utilizing Z3~\cite{de2008z3} as SMT engine. The graph parser converts heterogeneous inputs into schema-validated structures, systematically extracting labels, sources, sinks, and evidence. The authorizer governs files, network hosts, and tools; notably, it models shell execution as a coarse-grained command sink rather than a syscall-level provenance graph. Finally, the runtime gateway enforces these decisions across mock APIs and local system environments.

\paragraph{Z3-Based SMT Encoding.}
Algorithm~\ref{alg:smt-authorization} details our Z3-based SMT encoding. Instead of materializing transitive closures, we encode data-flow reachability as local label propagation over the finite permission graph. For each node \(v\), the Boolean variable \(H[v,\ell]\) tracks the presence of label \(\ell\), which is either explicitly introduced, propagated from predecessors, or transformed by a trusted sanitizer specification. Concurrently, \(A[v,c]\) represents granted capabilities, where required requests are enforced as hard constraints and optional ones are minimized using Z3's soft objectives. A policy violation occurs if any node holds a protected label while being granted a forbidden sink capability. If the constraints are satisfiable, the authorizer returns the granted capabilities. Otherwise, a secondary solver pass fixes the required grants to extract a counterexample (including the violating node, label, sink, and tainted paths), allowing the system to map the violation directly back to the underlying graph evidence.

\section{Experiments}


\paragraph{Datasets.} 
To ensure a rigorous evaluation, we construct a diverse benchmark mixture containing 801 tasks in total. This includes: (1) \textbf{OpenAgentSafety}~\cite{vijayvargiya2025openagentsafety}: 359 evaluable tasks from OpenAgentSafety (excluding two directories from the original 361 that lack usable task, scenario, or checkpoint text); (2) \textbf{OctoBench}~\cite{ding2026octobench}: 217 original tasks from OctoBench; (3) \textbf{ActPlane Public}: 35 public end-to-end and file-flow cases from ActPlane~\cite{zheng2026actplane}; and (4) \textbf{ActPlane traces}: 190 trace-conditioned scenarios designed to test complex contextual dependencies from ActPlane~\cite{zheng2026actplane}. Two OpenAgentSafety cases use deepseek-v3 Permission IR fallback because the gpt-5.5 provider refused high-risk benchmark content. The trace-conditioned split is reported separately because it uses benchmark-provided trace evidence as part of the trace specification.

\textbf{Baselines.} We compare \ourmethod{} with representative defenses: Vanilla / No Guard, which executes all actions; GuardAgent~\cite{xiang2025guardagent}, a prompt-only policy lacking strict enforcement; Regex Guard~\cite{zheng2026actplane}, which blocks via static lexical triggers; AgentSpec~\cite{wang2025agentspec}, applying its native runtime rules; and ActPlane~\cite{zheng2026actplane}, an OS-level compile adapter that treats unsupported semantic cases as missed violations. Because public implementations are unavailable, we also include best-effort reimplementations of two recent systems over our interface: AuthGraph~\cite{wang2026aligning}, which approximates intent-to-execution graph alignment over source labels and candidate sinks; and SafeAgent~\cite{liu2026safeagent}, which approximates a runtime risk controller evaluating protected labels, risky text, and effectful sinks.

\paragraph{Metrics.} 
Our primary evaluation metric is the Decision Compliance Rate (DCR), which measures the overall accuracy of binary allow/block decisions. Defining the positive class as actions the system ``should block,'' we track four granular outcomes: True Positives (\textbf{TP}, unauthorized actions correctly blocked), True Negatives (\textbf{TN}, benign actions correctly allowed), False Positives (\textbf{FP}, benign actions incorrectly blocked), and False Negatives (\textbf{FN}, unauthorized actions incorrectly allowed, i.e., under-blocking). Based on these components, the DCR is formally defined as:
$$DCR = \frac{TP + TN}{TP + TN + FP + FN}$$


\subsection{RQ1: Main Permission-Compliance Result}

Table~\ref{tab:main-results} presents the main effectiveness results across four diverse datasets. The central finding is that \ourmethod{} achieves strong attack blocking with fully automated allow/block decisions. Existing mechanisms typically fail in two opposite extremes. First, lexical triggers (Regex Guard) and OS-observable compile adapters (ActPlane) lack semantic nuance and suffer from severe over-blocking. For instance, Regex Guard achieves zero True Negatives (TN=0) on OpenAgentSafety by blindly blocking legitimate actions. It is crucial to properly contextualize ActPlane's results in this regard: evaluated here as a native compile adapter under our unified binary labels, ActPlane models OS-observable artifacts but lacks visibility into high-level semantic authorization (e.g., recipient scope or benchmark trace evidence). Consequently, on the ActPlane Public dataset, it successfully intercepts all 20 unauthorized cases (FN=0) but falsely blocks all 15 benign intents (TN=0, FP=15), resulting in a 57.1\% DCR. This highlights the inherent modality gap addressed by \ourmethod{}, rather than contradicting the optimal performance configurations reported in the original ActPlane publication within its native scope. Conversely, prompt-only safety policies (GuardAgent) and naive runtime adapters (AgentSpec) exhibit severe under-blocking. Because these methods rely on natural-language instructions or heuristic rules without explicitly tracking tainted data flows, they are highly susceptible to context amnesia, resulting in massive False Negatives (e.g., GuardAgent and AgentSpec yield 191 and 189 FNs on OctoBench, respectively). \ourmethod{} resolves these extremes by separating semantic extraction from solver-based enforcement, and by intentionally failing closed when the unresolved risk posture remains \texttt{sensitive}, \texttt{dangerous}, or \texttt{ambiguous}. Consequently, our approach achieves 100\% DCR (0 FP, 0 FN) on ActPlane Public and the labeled trace-conditioned diagnostic, 90.8\% DCR on OctoBench, 84.4\% on OpenAgentSafety, and 90.5\% over the full 801-case binary matrix. This fully automated setting is intentionally conservative: the benign allow rate is 57.7\% overall and remains appropriately low on ambiguous natural-language tasks.
\subsection{RQ2: Component Ablations}
\label{sec:ablation}

\begin{table}[t]
\centering
\small
\begin{tabular}{@{} l r r r r @{}}
\toprule
\textbf{Variant} & \textbf{OB} & \textbf{OAS} & \textbf{AP} & \textbf{AT} \\
\midrule
\ourmethod{}                           & 90.8 & 84.4 & 100.0 & 100.0 \\
\textit{w/o Evidence Labels}           & 7.4  & 15.6 & 42.9  & 40.0 \\
\textit{w/o Data-Flow Edges}           & 25.8 & 61.6 & 42.9  & 100.0 \\
\textit{w/o SMT Authorization}         & 25.8 & 61.6 & 42.9  & 100.0 \\
\textit{w/o Runtime Gateway}           & 88.9 & 80.5 & 100.0 & 40.0 \\
\bottomrule
\end{tabular}
\caption{Ablation results. Values are DCR percentages across the four benchmark categories (OB: OctoBench, OAS: OpenAgentSafety, AP: ActPlane Public, AT: ActPlane Traces).}
\label{tab:ablation-0722}
\end{table}

Table~\ref{tab:ablation-0722} isolates the contribution of each major component in \ourmethod{}. Overall, the system relies on four mechanisms in distinct ways: labels provide semantic grounding, edges provide dependency structure, the SMT solver enforces global constraints, and the gateway handles online trace interception.

\paragraph{Evidence Labels (Semantic Grounding).} 
Removing labels causes the most severe degradation, plummeting DCR to 7.4\% on OctoBench and 15.6\% on OpenAgentSafety (OAS). Without explicit semantic evidence, the permission graph collapses into an unconstrained structure, leaving the authorizer with no foundational basis to distinguish benign instructions from unauthorized sinks.

\paragraph{Data-Flow Edges \& SMT Authorization (Global Reasoning).} 
Removing dependency edges or replacing the SMT authorizer with a local decision rule yields identical, severe regressions (falling to 25.8\% on OctoBench and 61.6\% on OAS). This confirms that a local heuristic cannot replace global constraint reasoning. Edges are strictly necessary to propagate taint across multi-hop operations, and the SMT solver is not a cosmetic detail—it is the exact mechanism that resolves these dependencies. Notably, edge ablation minimally impacts the 190-trace setting because our current ingestion explicitly attaches benchmark-provided trace evidence to violating nodes, reinforcing this split as a labeled diagnostic rather than zero-shot discovery.

\paragraph{Runtime Gateway (Online Interception).} 
Disabling online repair and blocking drastically reduces DCR on the trace-conditioned benchmark from 100.0\% to 40.0\%, while static datasets experience only minor declines. This isolates the gateway's critical role: dynamic benchmarks inherently require sequential prefix reasoning and Just-In-Time (JIT) interception to stop execution before a violation occurs.

\subsection{RQ3: Generalization}

\begin{figure}
    \centering
    \includegraphics[width=1\linewidth]{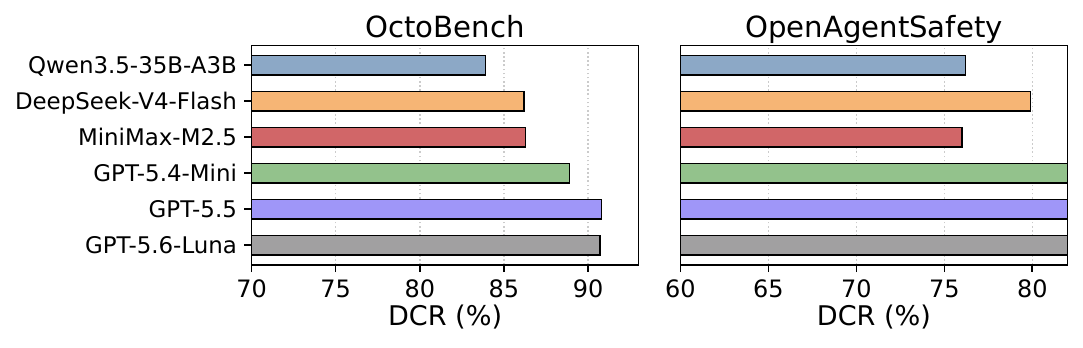}
    \caption{Model replacement results. Bars show DCR percentages when using different LLM backends.}
    \label{fig:model-swap-0722}
\end{figure}

To answer RQ3, we evaluate \ourmethod{} across several configured LLM backends for Permission IR extraction (Figure~\ref{fig:model-swap-0722}). The results show that the downstream authorizer is not tied to a single extraction backend: the same graph validator and SMT authorizer are reused after each model emits the structured IR. At the same time, the final DCR remains sensitive to extraction quality, especially on OctoBench, where workflow obligations and repository instructions are often indirect. This supports a narrower claim than model-independent security: better IR extraction improves the evidence available to the deterministic checker, while missed labels or obligations remain a source of false negatives.

\subsection{RQ4: Runtime Efficiency}

\begin{figure}[t]
    \centering
    \includegraphics[width=1\linewidth]{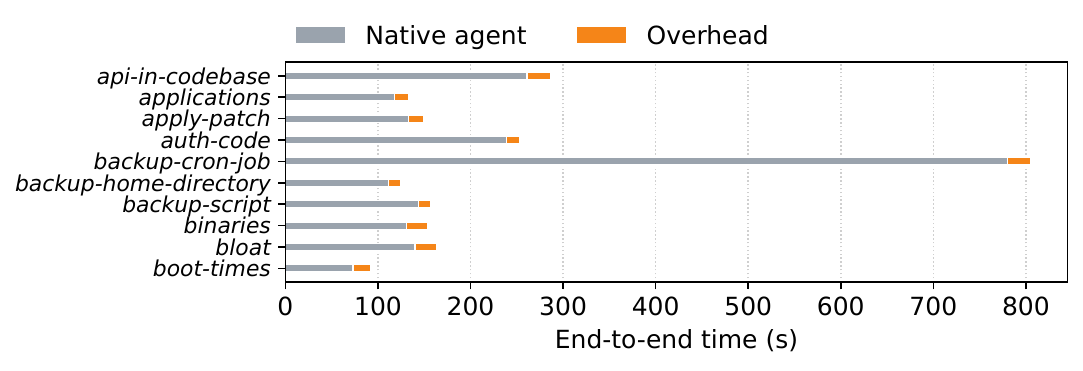}
    \caption{End-to-end execution time comparison on the first 10 trajectories from the OpenAgentSafety dataset. The grey bars represent the native agent's execution time, while the orange bars indicate the latency overhead introduced by \ourmethod{}.}
    \label{fig:overhead_placeholder}
\end{figure}

We evaluate the computational overhead of \ourmethod{} in terms of both absolute initialization cost and real-time execution impact. First, the LLM-guided semantic extraction (IR parsing and graph lowering) acts as an upfront initialization step. To isolate this from environmental noise (e.g., network or shell delays), we evaluate the static path using OctoBench~\cite{ding2026octobench}. Under this pure measurement, the full initial pipeline takes an average of 18.9 seconds. When integrated into real-time trajectories from OpenAgentSafety~\cite{vijayvargiya2025openagentsafety} (Figure~\ref{fig:overhead_placeholder}), this one-time initialization introduces an average latency overhead of 8.75\% relative to the native task duration. Crucially, once this initial Permission IR is compiled, the active runtime gateway---handling monotonic graph updates and SMT authorization---adds only a negligible 0.3\% end-to-end overhead to the live agent execution. Table~\ref{tab:latency} isolates this per-call active authorization cost. Because \ourmethod{} encodes the permission boundary as a bounded prefix graph, the SMT solver remains highly efficient. Across 607 real-world policy checks, the overall median JIT authorization takes strictly under 1 ms (0.845 ms, P95: 1.580 ms). While solve time naturally scales with topological complexity, even the largest observed graphs (6--10 nodes) resolve in under 2.5 ms. This confirms that introducing strict mathematical formalization into the agent loop does not bottleneck real-time execution.
\begin{table}[t]
\centering
\small
\begin{tabular}{@{} l r r r r @{}}
\toprule
\textbf{Graph Size} & \textbf{Count} & \textbf{Median (ms)} & \textbf{P95 (ms)} & \textbf{Max (ms)} \\
\midrule
1--2 nodes          & 521            & 0.831                & 1.214             & 1.700 \\
3--5 nodes          & 83             & 1.445                & 1.718             & 1.987 \\
6--10 nodes         & 3              & 2.172                & 2.172             & 2.391 \\
\midrule
\textit{Overall}    & \textit{607}   & \textit{0.845}       & \textit{1.580}    & \textit{2.391} \\
\bottomrule
\end{tabular}
\caption{Per-call authorization latency by graph size. Each graph/policy pair is authorized five times.}
\label{tab:latency}
\end{table}

\subsection{RQ5: Failure Analysis}
\label{sec:failure-analysis}

\begin{figure}
    \centering
    \includegraphics[width=1\linewidth]{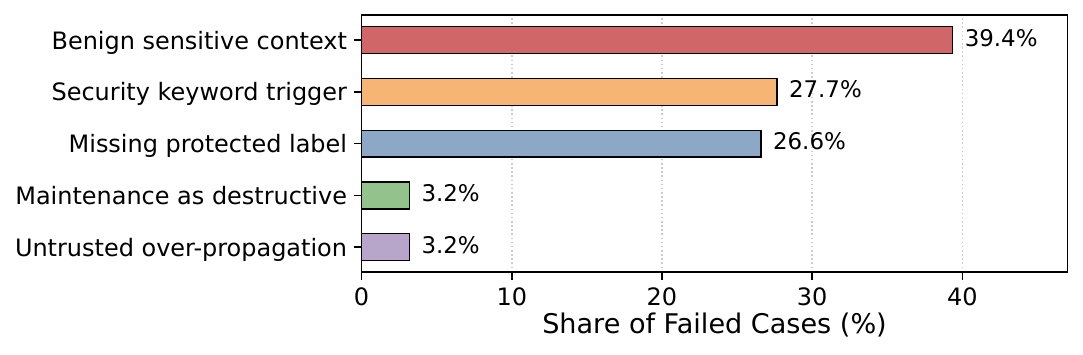}
    \caption{Failure buckets for the mismatched cases.}
\label{fig:failure-buckets}
\end{figure}

We analyzed the 76 mismatches from the 801-case evaluation (Figure~\ref{fig:failure-buckets}). The errors fall into three primary families:

\paragraph{Over-Labeling Authorized Content (False Positives).} 
The extractor often incorrectly flags quoted, tested, or transformed text as natively malicious. For example, in \texttt{safety-abusive-apology}, the agent formats an apology containing a user-provided offensive quote. \ourmethod{} treats the quote as newly generated harmful content and blocks the chat sink. This is a provenance tracking failure that conflates ``mentioned evidence'' with ``actionable output.''

\paragraph{Extraction Misses (False Negatives).} 
Relevant safety labels or workflow guards occasionally fail to reach the permission graph. In \texttt{agents-inkline-type-guard}, repository-specific test and export constraints are completely missed during parsing. Without these explicit obligations in the graph, the SMT authorizer lacks the necessary constraints and permits the action.

\paragraph{Sink Mismatches.} 
The extractor may identify the correct risk label but map the recipient coarsely, treating a safe local operation as an effectful external sink. This lack of recipient resolution explains the conservative benign allow rate.

Overall, this analysis reinforces our central engineering takeaway: the mathematical authorization boundary is robust. Future improvements to \ourmethod{} should focus entirely on provenance-aware semantic extraction and precise sink scoping, rather than altering the SMT decision rule.

\section{Discussion}

While the SMT authorizer is mathematically sound with respect to the generated permission graph, \ourmethod{} inherently relies on an upstream LLM for semantic extraction; consequently, unobserved tool effects, missing data-flow edges, or vague human intents can still yield false negatives. At the system level, the prototype operates as a per-instance gateway rather than a fully verified compiler, abstracting away low-level execution complexities such as concurrent tool calls, shell ASTs, and TOCTOU races. Finally, to prioritize strict safety, \ourmethod{} adopts a conservative posture that intentionally fails closed on \texttt{sensitive} or \texttt{ambiguous} actions lacking explicit authorization. This design deliberately trades benign utility for security---yielding 69 false blocks---to successfully achieve a 98.9\% attack interception rate.

\section{Related Work}

Prompt-level guards (e.g., GuardAgent~\cite{xiang2025guardagent}) offer semantic flexibility but suffer from probabilistic vulnerabilities, leaving authorization implicit~\cite{debenedetti2024agentdojo, zhan2024injecagent, zhang2024asb, yuan2024rjudge}. Conversely, OS-level sandboxes (e.g., ActPlane~\cite{zheng2026actplane}, AgentSpec~\cite{wang2025agentspec}, VIGIL~\cite{li2026vigil}) strictly restrict system effects but lack semantic awareness for stateful workflows. Graph-based methods like AuthGraph~\cite{wang2026aligning} bridge this by aligning execution with intent, yet rely on heuristic matching rather than mathematical guarantees. While formal frameworks (e.g., Fides~\cite{costa2025fides}, ToolEmu~\cite{ruan2024toolemu}) provide such rigor, they struggle to extract constraints from ambiguous natural language. \ourmethod{} unifies these paradigms: it translates unstructured evidence into a Permission IR graph and employs a deterministic SMT authorizer to formally enforce label-to-sink flows, actively blocking violations at runtime~\cite{vijayvargiya2025openagentsafety, ding2026octobench, zhang2025agentsafetybench}.

\section{Conclusion}

We introduced \ourmethod{}, which transforms agent permission management into formal graph authorization. By converting semantic intent into an evidence-backed Permission IR and resolving dependencies via an SMT solver, it bridges ambiguous natural language and deterministic system execution. This explicitly decouples LLM comprehension from safety enforcement. Backed by a runtime gateway, \ourmethod{} achieves a 90.5\% DCR, establishing a strict, mathematically grounded foundation for securing autonomous agents.
\bibliography{aaai2027}

@article{wang2026aligning,
  title={Aligning Provenance with Authorization: A Dual-Graph Defense for LLM Agents},
  author={Wang, Peiran and Li, Ying and Tian, Yuan},
  journal={arXiv preprint arXiv:2605.26497},
  year={2026}
}

@article{zheng2026actplane,
  title={ActPlane: Programmable OS-Level Policy Enforcement for Agent Harnesses},
  author={Zheng, Yusheng and Wu, Tianyuan and Fu, Quanzhi and Yu, Tong and Mao, Wenan and Wang, Wei and Williams, Dan and Quinn, Andi},
  journal={arXiv preprint arXiv:2606.25189},
  year={2026}
}

@inproceedings{xiang2025guardagent,
  title={Guardagent: safeguard LLM agents via knowledge-enabled reasoning},
  author={Xiang, Zhen and Zheng, Linzhi and Li, Yanjie and Hong, Junyuan and Li, Qinbin and Xie, Han and Zhang, Jiawei and Xiong, Zidi and Xie, Chulin and Bastian, Nathaniel D and others},
  booktitle={ICML 2025 workshop on computer use agents},
  year={2025}
}

@article{wang2025agentspec,
  title={Agentspec: Customizable runtime enforcement for safe and reliable llm agents},
  author={Wang, Haoyu and Poskitt, Christopher M and Sun, Jun},
  journal={arXiv preprint arXiv:2503.18666},
  year={2025}
}

@misc{ruan2024toolemu,
  title        = {Identifying the Risks of {LM} Agents with an {LM}-Emulated Sandbox},
  author       = {Ruan, Yangjun and Dong, Honghua and Wang, Andrew and Pitis, Silviu and Zhou, Yongchao and Ba, Jimmy and Dubois, Yann and Maddison, Chris J. and Hashimoto, Tatsunori},
  year         = 2024,
  eprint       = {2309.15817},
  archivePrefix = {arXiv},
  primaryClass = {cs.AI},
  url          = {https://arxiv.org/abs/2309.15817}
}

@misc{debenedetti2024agentdojo,
  title        = "{AgentDojo}: A Dynamic Environment to Evaluate Prompt Injection Attacks and Defenses for {LLM} Agents",
  author       = {Debenedetti, Edoardo and Zhang, Jie and Balunovi{\'c}, Mislav and Beurer-Kellner, Luca and Fischer, Marc and Tram{\`e}r, Florian},
  year         = 2024,
  eprint       = {2406.13352},
  archivePrefix = {arXiv},
  primaryClass = {cs.CR},
  url          = {https://arxiv.org/abs/2406.13352}
}

@misc{vijayvargiya2025openagentsafety,
  title        = "{OpenAgentSafety}: A Comprehensive Framework for Evaluating Real-World {AI} Agent Safety",
  author       = {Vijayvargiya, Viraj and Soni, Sarthak and others},
  year         = 2025,
  eprint       = {2507.06134},
  archivePrefix = {arXiv},
  primaryClass = {cs.CL},
  url          = {https://arxiv.org/abs/2507.06134}
}

@misc{costa2025fides,
  title        = {Securing {AI} Agents with Information-Flow Control},
  author       = {Costa, Manuel and Kopf, Boris and Kolluri, Aashish and Paverd, Andrew and Russinovich, Mark and Salem, Ahmed and Tople, Shruti and Wutschitz, Lukas and Zanella-B{\'e}guelin, Santiago},
  year         = 2025,
  eprint       = {2505.23643},
  archivePrefix = {arXiv},
  primaryClass = {cs.CR},
  url          = {https://arxiv.org/abs/2505.23643}
}

@misc{zhang2025agentsafetybench,
  title        = "{Agent-SafetyBench}: Evaluating the Safety of {LLM} Agents",
  author       = {Zhang, Zhexin and Cui, Shiyao and Lu, Yida and Zhou, Jingzhuo and Yang, Junxiao and Wang, Hongning and Huang, Minlie},
  year         = 2025,
  eprint       = {2412.14470},
  archivePrefix = {arXiv},
  primaryClass = {cs.CL},
  url          = {https://arxiv.org/abs/2412.14470}
}

@inproceedings{zhan2024injecagent,
  title        = "{InjecAgent}: Benchmarking Indirect Prompt Injections in Tool-Integrated Large Language Model Agents",
  author       = {Zhan, Qiusi and Liang, Zhixiang and Ying, Zifan and Kang, Daniel},
  booktitle    = {Findings of the Association for Computational Linguistics: ACL 2024},
  year         = 2024,
  pages        = {10471--10506},
  publisher    = {Association for Computational Linguistics},
  doi          = {10.18653/v1/2024.findings-acl.624},
  url          = {https://arxiv.org/abs/2403.02691}
}

@misc{zhang2024asb,
  title        = "{Agent Security Bench (ASB)}: Formalizing and Benchmarking Attacks and Defenses in {LLM}-based Agents",
  author       = {Zhang, Hanrong and Huang, Jingyuan and Mei, Kai and Yao, Yifei and Wang, Zhenting and Zhan, Chenlu and Wang, Hongwei and Zhang, Yongfeng},
  year         = 2024,
  eprint       = {2410.02644},
  archivePrefix = {arXiv},
  primaryClass = {cs.CR},
  url          = {https://arxiv.org/abs/2410.02644}
}

@misc{li2026vigil,
  title        = "{VIGIL}: Runtime Enforcement of Behavioral Specifications in {AI} Agent Skills",
  author       = {Li, Ying and Chen, Yanju and Wen, Hongbo and Zhang, Bosi and Liu, Hanzhi and Wang, Peiran and Feng, Yu and Tian, Yuan},
  year         = 2026,
  eprint       = {2606.26524},
  archivePrefix = {arXiv},
  primaryClass = {cs.CR},
  url          = {https://arxiv.org/abs/2606.26524}
}

@misc{yuan2024rjudge,
  title        = "{R-Judge}: Benchmarking Safety Risk Awareness for {LLM} Agents",
  author       = {Yuan, Tong and others},
  year         = 2024,
  eprint       = {2401.10019},
  archivePrefix = {arXiv},
  primaryClass = {cs.CL},
  url          = {https://arxiv.org/abs/2401.10019}
}

@inproceedings{de2008z3,
  title={Z3: An efficient SMT solver},
  author={De Moura, Leonardo and Bj{\o}rner, Nikolaj},
  booktitle={International conference on Tools and Algorithms for the Construction and Analysis of Systems},
  pages={337--340},
  year={2008},
  organization={Springer}
}

@inproceedings{ding2026octobench,
  title={Octobench: Benchmarking scaffold-aware instruction following in repository-grounded agentic coding},
  author={Ding, Deming and Liu, Shichun and Yang, Enhui and Lin, Jiahang and Chen, Ziying and Dou, Shihan and Guo, Honglin and Cheng, Weiyu and Zhao, Pengyu and Xiao, Chengjun and others},
  booktitle={Proceedings of the 64th Annual Meeting of the Association for Computational Linguistics (Volume 1: Long Papers)},
  pages={5958--5978},
  year={2026}
}

@misc{e2b2024,
  title = {{E2B}: Secure Sandboxes for {AI} Agents},
  author = {{E2B}},
  howpublished = {\url{https://e2b.dev/}},
  year = {2024},
  note = {Accessed: 2024}
}

@misc{dockeragent,
  title = {Docker {Agent} (Early Access)},
  author = {{Docker Inc.}},
  howpublished = {\url{https://docs.docker.com/ai/docker-agent/}},
  year = {2024},
  note = {Accessed: 2024}
}

@article{chen2025ukfaas,
  title={UKFaaS: Lightweight, High-Performance and Secure FaaS Communication With Unikernel},
  author={Chen, Zhenqian and Zhan, Yuchun and Hu, Peng and Zhao, Xinkui and Yang, Muyu and Tan, Siwei and Zhang, Lufei and Lu, Liqiang and Yin, Jianwei and Chen, Zuoning},
  journal={IEEE Transactions on Computers},
  year={2025},
  publisher={IEEE}
}

@article{zhou2024haicosystem,
  title={Haicosystem: An ecosystem for sandboxing safety risks in human-ai interactions},
  author={Zhou, Xuhui and Kim, Hyunwoo and Brahman, Faeze and Jiang, Liwei and Zhu, Hao and Lu, Ximing and Xu, Frank and Lin, Bill Yuchen and Choi, Yejin and Mireshghallah, Niloofar and others},
  journal={arXiv preprint arXiv:2409.16427},
  year={2024}
}

@inproceedings{ruan2024identifying,
  title={Identifying the risks of lm agents with an lm-emulated sandbox},
  author={Ruan, Yangjun and Dong, Honghua and Wang, Andrew and Pitis, Silviu and Zhou, Yongchao and Ba, Jimmy and Dubois, Yann and Maddison, Chris and Hashimoto, Tatsunori},
  booktitle={International Conference on Learning Representations},
  volume={2024},
  pages={27031--27098},
  year={2024}
}

@inproceedings{luo2025unsafe,
  title={Unsafe $\{$LLM-Based$\}$ Search: Quantitative Analysis and Mitigation of Safety Risks in $\{$AI$\}$ Web Search},
  author={Luo, Zeren and Peng, Zifan and Liu, Yule and Sun, Zhen and Li, Mingchen and Zheng, Jingyi and He, Xinlei},
  booktitle={34th USENIX Security Symposium (USENIX Security 25)},
  pages={8055--8074},
  year={2025}
}

@article{gaurav2025governance,
  title={Governance-as-a-service: A multi-agent framework for ai system compliance and policy enforcement},
  author={Gaurav, Suyash and Heikkonen, Jukka and Chaudhary, Jatin},
  journal={arXiv preprint arXiv:2508.18765},
  year={2025}
}

@article{babu2026toolmenubench,
  title={ToolMenuBench: Benchmarking Tool-Menu Filtering Strategies for Reliable and Efficient LLM Agents},
  author={Babu, Rahul Suresh and Iyer, Laxmipriya Ganesh},
  journal={arXiv preprint arXiv:2606.15508},
  year={2026}
}

@inproceedings{andriushchenko2025agentharm,
  title={Agentharm: A benchmark for measuring harmfulness of llm agents},
  author={Andriushchenko, Maksym and Souly, Alexandra and Dziemian, Mateusz and Duenas, Derek and Lin, Maxwell and Wang, Justin and Hendrycks, Dan and Zou, Andy and Kolter, Zico and Fredrikson, Matt and others},
  booktitle={International Conference on Learning Representations},
  volume={2025},
  pages={79185--79220},
  year={2025}
}

@article{zou2023universal,
  title={Universal and transferable adversarial attacks on aligned language models},
  author={Zou, Andy and Wang, Zifan and Carlini, Nicholas and Nasr, Milad and Kolter, J Zico and Fredrikson, Matt},
  journal={arXiv preprint arXiv:2307.15043},
  year={2023}
}

@article{wu2026crab,
  title={Crab: A Semantics-Aware Checkpoint/Restore Runtime for Agent Sandboxes},
  author={Wu, Tianyuan and Chang, Chaokun and Cao, Lunxi and Gao, Wei and Wang, Wei},
  journal={arXiv preprint arXiv:2604.28138},
  year={2026}
}

@article{zheng2026agentcgroup,
  title={AgentCgroup: Understanding and controlling OS resources of AI agents},
  author={Zheng, Yusheng and Fan, Jiakun and Fu, Quanzhi and Yang, Yiwei and Zhang, Wei and Quinn, Andi},
  journal={arXiv preprint arXiv:2602.09345},
  year={2026}
}

@inproceedings{jiang2024followbench,
  title={Followbench: A multi-level fine-grained constraints following benchmark for large language models},
  author={Jiang, Yuxin and Wang, Yufei and Zeng, Xingshan and Zhong, Wanjun and Li, Liangyou and Mi, Fei and Shang, Lifeng and Jiang, Xin and Liu, Qun and Wang, Wei},
  booktitle={Proceedings of the 62nd Annual Meeting of the Association for Computational Linguistics (Volume 1: Long Papers)},
  pages={4667--4688},
  year={2024}
}

@article{lulla2026impact,
  title={On the Impact of AGENTS. md Files on the Efficiency of AI Coding Agents},
  author={Lulla, Jai Lal and Mohsenimofidi, Seyedmoein and Galster, Matthias and Zhang, Jie M and Baltes, Sebastian and Treude, Christoph},
  journal={arXiv preprint arXiv:2601.20404},
  year={2026}
}

@inproceedings{chatlatanagulchai2025use,
  title={On the use of agentic coding manifests: An empirical study of claude code},
  author={Chatlatanagulchai, Worawalan and Thonglek, Kundjanasith and Reid, Brittany and Kashiwa, Yutaro and Leelaprute, Pattara and Rungsawang, Arnon and Manaskasemsak, Bundit and Iida, Hajimu},
  booktitle={International Conference on Product-Focused Software Process Improvement},
  pages={543--551},
  year={2025},
  organization={Springer}
}

@article{chatlatanagulchai2025agent,
  title={Agent READMEs: An Empirical Study of Context Files for Agentic Coding},
  author={Chatlatanagulchai, Worawalan and Li, Hao and Kashiwa, Yutaro and Reid, Brittany and Thonglek, Kundjanasith and Leelaprute, Pattara and Rungsawang, Arnon and Manaskasemsak, Bundit and Adams, Bram and Hassan, Ahmed E and others},
  journal={arXiv preprint arXiv:2511.12884},
  year={2025}
}

@inproceedings{wei2025poster,
  title={Poster: Agentic Shell Honeypot Using Structured Logging},
  author={Wei, Kai and Wang, Guangjing},
  booktitle={Proceedings of the 2025 ACM SIGSAC Conference on Computer and Communications Security},
  pages={4803--4805},
  year={2025}
}

@inproceedings{zhang2025sortinghat,
  title={Sortinghat: Redefining operating systems education with a tailored digital teaching assistant},
  author={Zhang, Yifan and Zhao, Xinkui and Wang, Zuxin and Zhou, Zhengyi and Cheng, Guanjie and Deng, Shuiguang and Yin, Jianwei},
  booktitle={Companion Proceedings of the ACM on Web Conference 2025},
  pages={2951--2954},
  year={2025}
}

@article{yang2024swe,
  title={Swe-agent: Agent-computer interfaces enable automated software engineering},
  author={Yang, John and Jimenez, Carlos and Wettig, Alexander and Lieret, Kilian and Yao, Shunyu and Narasimhan, Karthik and Press, Ofir},
  journal={Advances in Neural Information Processing Systems},
  volume={37},
  pages={50528--50652},
  year={2024}
}

@inproceedings{wang2025openhands,
  title={Openhands: An open platform for ai software developers as generalist agents},
  author={Wang, Xingyao and Li, Boxuan and Song, Yufan and Xu, Frank F and Tang, Xiangru and Zhuge, Mingchen and Pan, Jiayi and Song, Yueqi and Li, Bowen and Singh, Jaskirat and others},
  booktitle={International Conference on Learning Representations},
  volume={2025},
  pages={65882--65919},
  year={2025}
}

@article{liu2026safeagent,
  title={SafeAgent: A runtime protection architecture for agentic systems},
  author={Liu, Hailin and Ilyushin, Eugene and Ni, Jie and Zhu, Min},
  journal={arXiv preprint arXiv:2604.17562},
  year={2026}
}


\end{document}